\documentclass[runningheads]{llncs}
\usepackage[T1]{fontenc}
\usepackage{graphicx}
\usepackage{amsmath}
\usepackage{amssymb}
\usepackage[colorinlistoftodos]{todonotes}
\presetkeys{todonotes}{inline}{}
\usepackage{booktabs}
\usepackage{subcaption}
\usepackage{multirow}
\usepackage{tabularx}
\usepackage[htt]{hyphenat} 
\usepackage{float}
\usepackage{placeins}
\usepackage{wrapfig}
\begin{document}
\title{Utilizing Metadata for Better Retrieval-Augmented Generation}
%

\author {Raquib Bin Yousuf\inst{1}\orcidID{0009-0001-1245-0450} \and Shengzhe Xu\inst{1}\orcidID{0009-0008-8589-6526} \and Mandar Sharma\inst{1}\orcidID{0000-0002-7012-9323} \and Andrew Neeser\inst{1} \and Chris Latimer\inst{2} \and Naren Ramakrishnan\inst{1}\orcidID{0000-0002-1821-9743}}
\authorrunning{Yousuf et al.}
\institute{Virginia Tech, Virginia, USA \\
\email{\{raquib, shengzx, mandarsharma, aneeser24, naren\}@vt.edu}
\and Vectorize.io, Colorado, USA \\
\email{chris.latimer@vectorize.io}}

\maketitle              
\begin{abstract}
Retrieval-Augmented Generation systems depend on retrieving semantically relevant document chunks to support accurate, grounded outputs from large language models. In structured and repetitive corpora such as regulatory filings, chunk similarity alone often fails to distinguish between documents with overlapping language. Practitioners often flatten metadata into input text as a heuristic, but the impact and trade-offs of this practice remain poorly understood.  
We present a systematic study of metadata-aware retrieval strategies, comparing plain-text baselines with approaches that embed metadata directly. Our evaluation spans metadata-as-text (prefix and suffix), a dual-encoder unified embedding that fuses metadata and content in a single index, dual-encoder late-fusion retrieval, and metadata-aware query reformulation. Across multiple retrieval metrics and question types, we find that prefixing and unified embeddings consistently outperform plain-text baselines, with the unified at times exceeding prefixing while being easier to maintain. 
Beyond empirical comparisons, we analyze embedding space, showing that metadata integration improves effectiveness by increasing intra-document cohesion, reducing inter-document confusion, and widening the separation between relevant and irrelevant chunks. Field-level ablations show that structural cues provide strong disambiguating signals.
Our code, evaluation framework, and the \textsc{RAGMATE-10K} dataset are publicly hosted\footnote{https://github.com/raquibvt/RAGMate}.
\keywords{Retrieval-Augmented Generation (RAG) \and Metadata-aware Retrieval \and Dense Retrieval \and Query Reformulation \and Benchmark Datasets}
\end{abstract}
\begin{figure}[t]
    \centering
    \begin{subfigure}{0.48\linewidth}
        \centering
        \includegraphics[width=\linewidth]{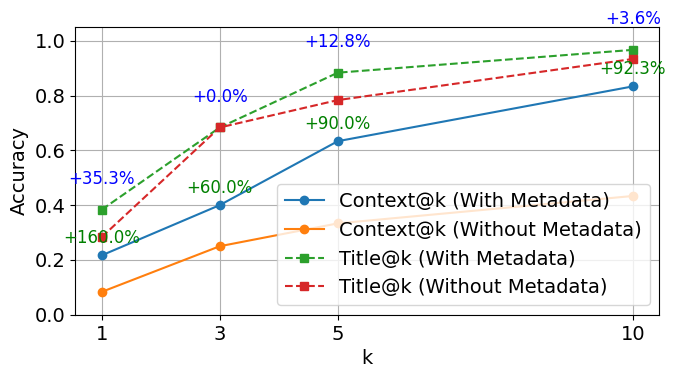}
        \caption{General questions}
    \end{subfigure}
    \hfill
    \begin{subfigure}{0.48\linewidth}
        \centering
        \includegraphics[width=\linewidth]{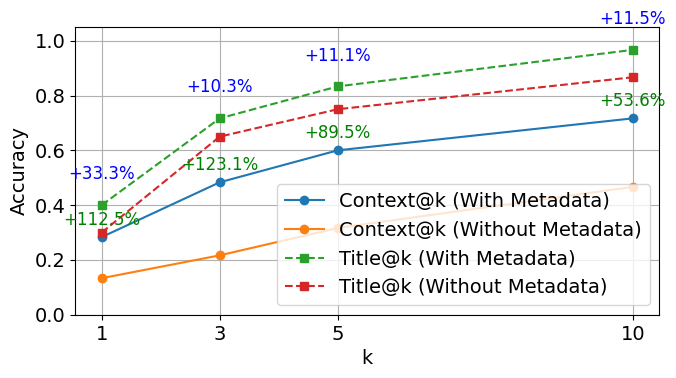}
        \caption{In-depth questions}
    \end{subfigure}
    \caption{Context@K and Title@K metric improvement in retrieval when using and not using metadata across query types (Dual Encoder Unified Embeddings)}
    \label{fig:uni_findings_1}
\end{figure}
\section{Introduction}
Large Language Models (LLMs) have become central to information access, enabling systems that answer questions, summarize documents, and reason over unstructured corpora~\cite{touvron2023llama,achiam2023gpt}. Retrieval-Augmented Generation (RAG) architectures extend these models by first retrieving relevant context and then conditioning generation on it~\cite{lewis2020retrieval}. This hybrid setup has become the dominant paradigm for open-domain question answering, long-document understanding, and domain-specific inference~\cite{gao2023retrieval,cuconasu2024power}.

However, RAG pipelines are only as reliable as the context they retrieve. When document corpora are repetitive or semantically ambiguous, such as regulatory filings, legal records, or scientific papers, retrievers often return superficially relevant but substantively unhelpful chunks~\cite{gao2023retrieval,ryu2023retrieval}. In these settings, small differences in document structure or entity context (for example company name, fiscal year, or section title) become essential for retrieving the correct information.

A particularly illustrative case is the U.S. Securities and Exchange Commission (SEC) Form 10-K filings. These documents follow rigid templates, reuse language across companies and years, and contain subtle, structure-dependent variations. Such properties challenge traditional vector-based retrieval, which relies on chunk-level semantic embeddings~\cite{devlin2019bert,gao2023retrieval} that often lack sufficient discriminative power. Many of the disambiguating signals already exist in structured metadata such as filing year, form type, section heading, and industry classification, yet these fields are usually applied only for post-retrieval filtering~\cite{zhou2022least}.

For example, consider the query ``What risks does the company identify related to supply chain disruptions?'' Without metadata, many chunks across different companies, years, and sections contain nearly identical language, causing semantic-only retrievers to return plausible but incorrect context. Metadata such as company, fiscal year, and section title disambiguate the query intent, anchoring retrieval to the correct filing and section.

SEC filings have therefore become a common testbed for financial QA with LLMs~\cite{islam2023financebench,lai2024sec,yoash2025secque}, alongside broader benchmarks in finance~\cite{xie2024finben,chen2021finqa}. Together, these studies highlight both the opportunities and the limitations of RAG pipelines on repetitive, domain-specific corpora. 
Motivated by recent work in contextual fine-tuning and long-context prompting~\cite{anthropic_contextual,asai2024self}, we ask:
\begin{quote}
\emph{Can metadata be utilized as a first-class input in RAG pipelines, not just as a filter, but as embedded content that improves retrieval quality and downstream answer correctness?}
\end{quote}
We investigate whether metadata-aware retrieval benefits from more modular designs such as dual encoders that embed content and metadata separately, or query reformulation that explicitly surfaces metadata cues. As a simple baseline, we also evaluate adding metadata directly to the chunk as text. In particular, prefixing metadata before the chunk yields strong retrieval accuracy, though it is computationally expensive since any metadata update requires re-embedding the full index.
\begin{figure}[t]
    \centering
    \begin{subfigure}{0.45\linewidth}
        \centering
        \includegraphics[width=\linewidth]{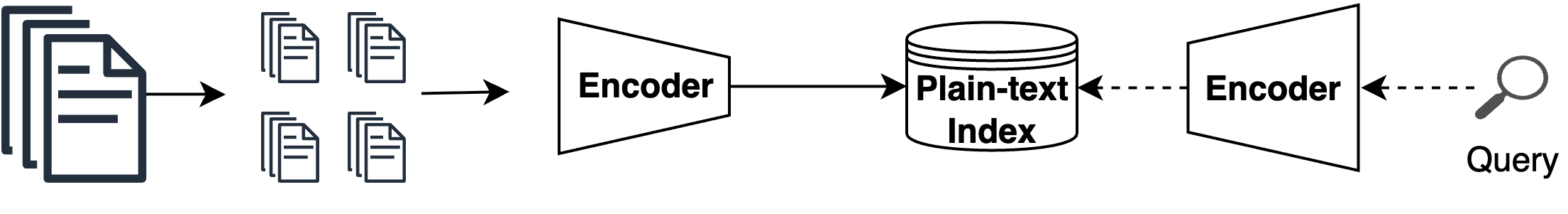}
        \caption{Plain text index}
    \end{subfigure}
    \hfill
    \begin{subfigure}{0.45\linewidth}
        \centering
        \includegraphics[width=\linewidth]{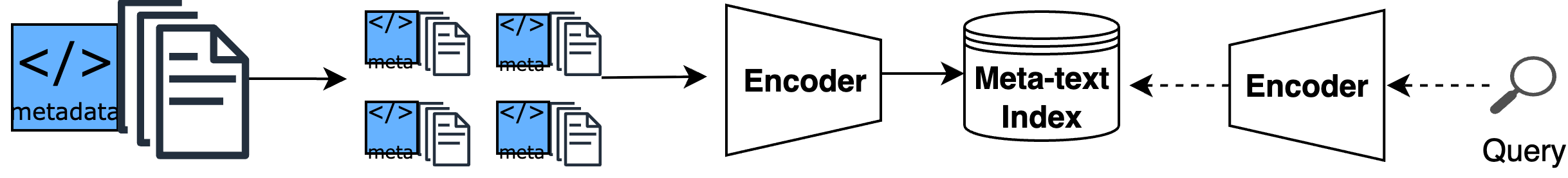}
        \caption{Metadata as part of text}
    \end{subfigure}
    \begin{subfigure}{0.46\linewidth}
        \centering
        \includegraphics[width=\linewidth]{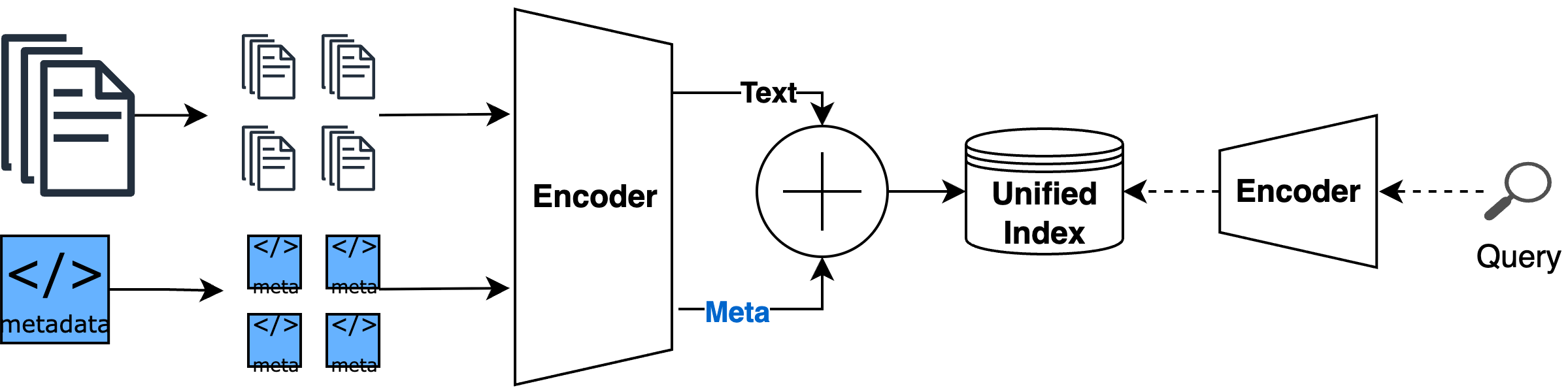}
        \caption{Dual encoder (unified index)}
    \end{subfigure}
    \hfill
    \begin{subfigure}{0.46\linewidth}
        \centering
        \includegraphics[width=\linewidth]{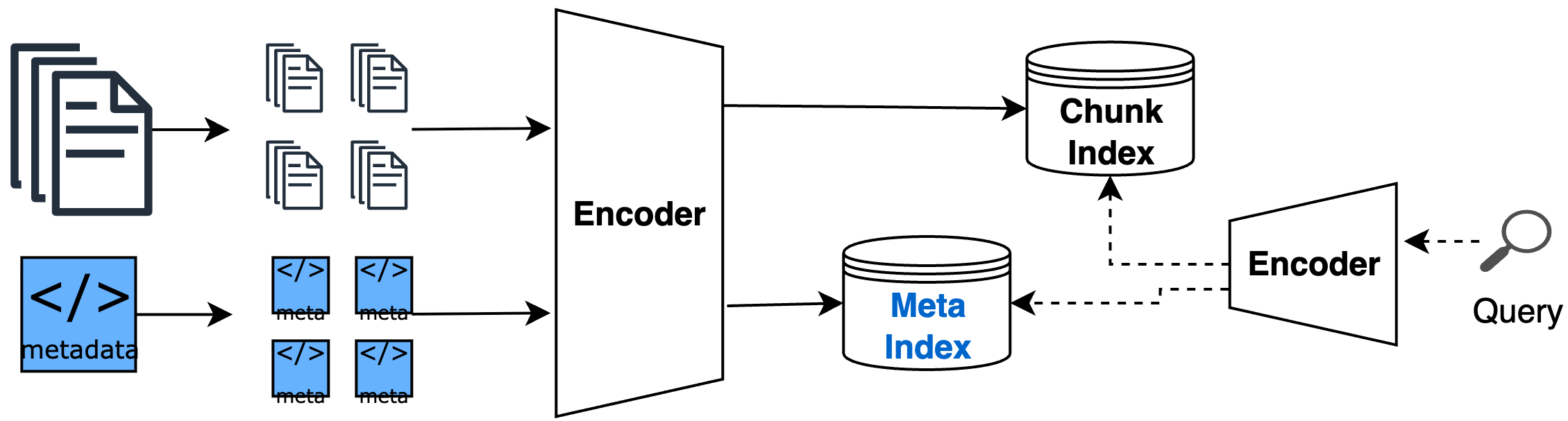}
        \caption{Dual encoder (late fusion)}
    \end{subfigure}
    \vskip\baselineskip
 \caption{Conceptual overview of document and metadata embedding strategies for retrieval. Dotted arrows indicate query propagation.}
    \label{fig:conceptual_overview}
\end{figure}
To make maintenance lighter, we propose a dual-encoder framework in which metadata and content are embedded independently and then combined. Metadata embeddings only need to be computed once per field and can be fused with precomputed text embeddings, avoiding costly re-embedding of the full corpus. Among these variants, unified embeddings, where metadata and content vectors are summed into a single index, achieve accuracy comparable to, and sometimes surpassing, prefixing while simplifying serving. In contrast, late-fusion scoring is less competitive.

Beyond empirical comparisons, we also analyze why metadata integration improves retrieval. Through statistical analysis of embedding spaces, we show that metadata increases intra-document cohesion, decreases inter-document confusion, and expands the variance of similarity scores, creating a more discriminative geometry for retrieval. This analysis confirms that metadata does not simply add tokens but reshapes the embedding landscape in a way that improves separability between relevant and irrelevant chunks.

We summarize our contributions as follows:
\begin{itemize}
    \item To our knowledge, the first systematic study showing that metadata signals significantly improve retrieval on RAG systems.  
    \item A dual-encoder framework that embeds metadata and content separately, improving index maintenance while achieving accuracy on par with, and sometimes better than, prefixing.
    \item A statistical embedding-space analysis that explains these gains, demonstrating increased intra-document cohesion, reduced inter-document confusion, and clearer separation between relevant and irrelevant chunks.  
    \item Field-level ablations show that company and year fields provide strong disambiguating signals, while section titles offer more modest gains.
    \item Release of \textsc{RAGMATE-10K}, a metadata-rich benchmark with chunk grounded QA for reproducible evaluation.  
\end{itemize}
We review related work (§\ref{sec:related}), introduce metadata-as-text and dual-encoder designs (§\ref{ssec:dual}), describe the \textsc{RAGMATE-10K} dataset and setup (§\ref{sec:methods}), present results (§\ref{sec:findings}), analyze embedding geometry (§\ref{sec:embedding_analysis}), report ablations on section titles (§\ref{sec:context}), and conclude with limitations and future work.

\section{Related Work}
\label{sec:related}
Information access has long motivated technologies from search engines~\cite{page1999pagerank,brin1998anatomy} to diverse downstream applications~\cite{hu2008collaborative,li2010contextual}. The advent of attention-based models~\cite{vaswani2017attention,devlin2019bert} and modern chat-based LLMs~\cite{touvron2023llama,achiam2023gpt} has made retrieval-augmented generation (RAG)~\cite{lewis2020retrieval} a central paradigm. While standalone LLMs often hallucinate, which is problematic in high-stakes domains such as medicine~\cite{zakka2024almanac}, law~\cite{ryu2023retrieval}, intelligence analysis~\cite{yousuf2024llm}, and software engineering~\cite{Lee2025LLMSpreadsheet}, RAG mitigates this by grounding generation in external knowledge. Approaches span both fine-tuned retrievers~\cite{lin2023ra,asai2024self} and retraining-free pipelines that dominate current practice~\cite{gao2023retrieval,cuconasu2024power}.  

Within retrieval itself, methods differ in how they represent knowledge: some embed raw document chunks~\cite{gao2023retrieval}, others use (question, answer) pairs~\cite{zhou2022least}, and some generate synthetic queries or summaries to improve matching~\cite{gao2023precise,baek2023knowledge,luo2023reasoning}. These span dense embedding approaches and sparse, keyword-based retrieval. Recent work such as EnrichIndex~\cite{chen2025enrichindex} further explores offline LLM-based enrichment by generating summaries, purposes, and synthetic QA pairs to strengthen first-stage retrieval without online LLM costs.

Most prior work leverages metadata only indirectly, either for filtering or through late-fusion signals. Financial QA benchmarks such as FinanceBench~\cite{islam2023financebench}, SEC-QA~\cite{lai2024sec}, and SecQue~\cite{yoash2025secque}, along with broader datasets in finance~\cite{xie2024finben,chen2021finqa}, highlight the challenges of repetitive, domain-specific corpora but stop short of systematically evaluating metadata integration. While some recent approaches enrich documents with additional LLM-generated semantic context, they do not isolate the role of structured metadata as a retrieval signal. Our work differs by treating metadata as a first-class retrieval signal, comparing simple metadata-as-text strategies with modular dual-encoder designs.  

\section{Metadata as a First-Class Signal}
We begin with the simplest approach to metadata integration: concatenating structured metadata fields directly to the chunk text before embedding. This technique treats metadata as part of the semantic input (\emph{metadata-as-text}, MaT) and encodes it into the same vector space as content. 

\subsection{Metadata as Text: A Minimal Baseline}
\label{ssec:base}
Let the corpus be a set of \(N\) document chunks with associated structured metadata:
\[
D \;=\; \{(m_i, c_i)\}_{i=1}^N,
\]
where \(c_i\) is the text of the \(i\)-th chunk, and \(m_i\) is a key–value map of structured fields (e.g., company, form\_type, section, year).

We define a serialization function that produces a compact, human-readable header from the metadata:
\[
s(m_i) \;=\; ``company: \{...\}; form: \{...\}; section: \{...\}; year: \{...\}''.
\]
Using this header, we construct a metadata-prefixed chunk string via a concatenation operator:
\[
\tilde{c}_i \;=\; \texttt{concat}\big(s(m_i),\, c_i\big).
\]
We also evaluate a suffix variant that appends the metadata header after the chunk:
\(\tilde{c}_i^{\text{suf}}=\texttt{concat}\big(c_i,\, s(m_i)\big)\).
We report results for both prefix
\((\tilde{c}_i^{\text{pre}}=\texttt{concat}\big(s(m_i),\, c_i\big))\)
and suffix in Table~\ref{tab:retrieval_summary}, while figures use the prefix variant unless explicitly noted.

An off-the-shelf text encoder \(f_{\theta}\) (frozen) maps strings to \(d\)-dimensional vectors:
\[
\tilde{\mathbf{e}}_i \;=\; f_{\theta}(\tilde{c}_i) \in \mathbb{R}^d.
\]
Given a user query \(q\), we compute its embedding \(\mathbf{e}_q = f_{\theta}(q)\) and retrieve by cosine similarity over the single MaT index:
\[
\text{Score}_{\text{MaT}}(q, i) \;=\; \cos\!\big(\mathbf{e}_q,\, \tilde{\mathbf{e}}_i\big),
\quad
\text{rank by } \text{Score}_{\text{MaT}}.
\]

\subsection{Dual Encoders: Modular Integration}
\label{ssec:dual}
Flattening metadata into the chunk text (Section~\ref{ssec:base}) improves retrieval but is computationally expensive, since any metadata update requires re-embedding the full chunk index. To address this, we design dual-encoder approaches that embed content and metadata separately, making updates lighter and more modular. Within this framework, we first present a unified single-index that merges both signals directly in embedding space, retaining the simplicity of serving while avoiding costly re-indexing. We then contrast it with a late-fusion dual encoder that combines scores at query time, and finally describe query-side strategies that surface metadata cues in the query.

\subsubsection{Unified Single-Index via Weighted-Sum Fusion}
\label{sssec:sumfusion}
Let the corpus be $D=\{(m_i,c_i)\}_{i=1}^N$, where $c_i$ is chunk text and $m_i$ is a key--value metadata map (e.g., \texttt{company}, \texttt{form}, \texttt{year}, \texttt{section}). We encode content and metadata into the same $d$-dimensional space:
\[
\mathbf{e}^{\text{text}}_i = f^{\text{text}}_{\theta}(c_i), \qquad
\mathbf{e}^{\text{meta}}_i = f^{\text{meta}}_{\theta}(m_i).
\]
We L2-normalize both vectors and form a convex combination to build a single fused index:
\begin{align}
\label{eq:sumfusion_doc}
\hat{\mathbf{e}}^{\text{text}}_i &= 
   \frac{\mathbf{e}^{\text{text}}_i}{\|\mathbf{e}^{\text{text}}_i\|_2}, 
&\quad
\hat{\mathbf{e}}^{\text{meta}}_i &= 
   \frac{\mathbf{e}^{\text{meta}}_i}{\|\mathbf{e}^{\text{meta}}_i\|_2}, \nonumber \\[6pt]
\mathbf{e}^{\text{sum}}_i(\alpha) &= 
   \frac{\alpha\,\hat{\mathbf{e}}^{\text{text}}_i+(1-\alpha)\,\hat{\mathbf{e}}^{\text{meta}}_i}
        {\left\|\alpha\,\hat{\mathbf{e}}^{\text{text}}_i+(1-\alpha)\,\hat{\mathbf{e}}^{\text{meta}}_i\right\|_2},
&\quad \alpha &\in [0,1].
\end{align}
At query time, we embed the query once with the text encoder and retrieve by cosine similarity against the fused index. Since document embeddings are already L2-normalized, leaving the query unnormalized does not affect ranking under cosine similarity:
\begin{equation}
\label{eq:sumfusion_score}
\text{Score}_{\text{sum}}(q,i;\alpha)
\;=\;
\cos\!\big(\mathbf{e}^{\text{text}}_q,\ \mathbf{e}^{\text{sum}}_i(\alpha)\big).
\end{equation}
For inner-product distance, both $\mathbf{e}^{\text{text}}_q$ and $\mathbf{e}^{\text{sum}}_i(\alpha)$ should be L2-normalized to emulate cosine.

Eqs.~\eqref{eq:sumfusion_doc}--\eqref{eq:sumfusion_score} yield a single index of dimension $d$, without doubling as in concatenation. They keep metadata and text separable until fusion and avoid runtime score fusion.

\subsubsection{Dual Encoder with Late-Fusion Scoring}
\label{sec:latefusion}
Alternatively, we maintain two indices (content and metadata) and combine scores at query time. Late fusion exposes $\alpha$ at query time, which is useful for diagnostics, but requires two lookups. Given a query text embedding $\mathbf{e}^{\text{text}}_q$ and document embeddings $(\mathbf{e}^{\text{text}}_i, \mathbf{e}^{\text{meta}}_i)$, we compute
\begin{equation}
\label{eq:latefusion}
\text{Score}_{\text{late}}(q,i;\alpha)
= (1-\alpha)\,\cos\!\big(\mathbf{e}^{\text{text}}_q,\mathbf{e}^{\text{text}}_i\big)
 \;+\; \alpha\,\cos\!\big(\mathbf{e}^{\text{text}}_q,\mathbf{e}^{\text{meta}}_i\big),
\quad \alpha\!\in\![0,1].
\end{equation}
\subsection{Query-Side Strategies: Metadata-aware Reformulation}
\label{sec:queryside}
We apply an LLM-based reformulation operator to incorporate schema cues (e.g., company, form, year, section) into the query, which is then embedded with the text encoder:
\[
\phi_{\text{text}}(q)=\texttt{Reformulate}(q).
\]
This reformulated query can be used with both metadata-as-text retrieval and dual-encoder variants but it adds extra overhead during query time. Query reformulation is implemented as a single LLM call conditioned on the metadata schema (Table \ref{tab:metadata_schema}) and a small set of example values per field. Given a query, the LLM extracts explicit metadata constraints and rewrites the query to incorporate them. Retrieval then uses the rewritten query embedded with the same frozen text encoder as all other methods, leaving the retrieval pipeline unchanged.

\subsection{Embedding-Space Theory of Metadata Integration}
\label{sec:embedding_theory}
We consider a metadata-informed embedding $\tilde{\mathbf{e}}^{\star}_i$ that augments a chunk embedding $\mathbf{e}_i=f_\theta(c_i)$ with structured metadata, either through token-level prefixing (MaT) or vector-level fusion (Unified, see Sec.~\ref{sssec:sumfusion}). Let $d \in \mathcal{D}$ denote a document (e.g., a company–year SEC filing) and $i \in d$ a chunk belonging to $d$. The following propositions describe how such embeddings reshape the similarity landscape.
\begin{proposition}[Intra-document cohesion increases]
\label{prop:intra}
\[
\mathbb{E}_{i,j \in d}\!\left[\cos(\tilde{\mathbf{e}}_i,\tilde{\mathbf{e}}_j)\right]
\;>\;
\mathbb{E}_{i,j \in d}\!\left[\cos(\mathbf{e}_i,\mathbf{e}_j)\right].
\]
\end{proposition}
Metadata anchors chunks to their document identity, pulling them closer in embedding space.
\begin{proposition}[Inter-document confusion decreases]
\label{prop:inter}
\[
\mathbb{E}_{i \in d_1,\, j \in d_2,\, d_1 \neq d_2}\!\left[\cos(\tilde{\mathbf{e}}_i,\tilde{\mathbf{e}}_j)\right]
\;<\;
\mathbb{E}_{i \in d_1,\, j \in d_2,\, d_1 \neq d_2}\!\left[\cos(\mathbf{e}_i,\mathbf{e}_j)\right].
\]
\end{proposition}
Metadata provides discriminative cues (company, year, section) that reduce spurious similarity across different documents.
\begin{proposition}[Score variance increases]
\label{prop:variance}
\[
\mathrm{Var}\!\left[\cos(\mathbf{e}_q,\tilde{\mathbf{e}}_i)\right]
\;>\;
\mathrm{Var}\!\left[\cos(\mathbf{e}_q,\mathbf{e}_i)\right],
\qquad q \sim \text{typical queries}.
\]
\end{proposition}
Unified embeddings interpolate between content-only and metadata-only signals via a convex weight $\alpha$, and therefore inherit the above properties whenever $\alpha<1$. MaT achieves a similar effect through token-level injection, while Unified does so through vector-level fusion with tunable weighting. In effect, this creates a more structured space with clearer separation between relevant and irrelevant candidates.

\section{Methodology}
\label{sec:methods}
\subsection{Dataset: \textsc{RAGMATE-10K}}
We introduce \textsc{RAGMATE-10K}, a dataset of SEC 10-K filings designed to evaluate metadata-aware retrieval. It consists of 25 filings from five U.S. technology companies (Apple, Alphabet, Adobe, Oracle, Nvidia), each segmented into non-overlapping 350-token chunks with 50-token overlap. This yields \( N = 4{,}490 \) retrieval units, each represented as a tuple \( (m_i, c_i) \) where \( c_i \) is the text content and \( m_i \) its structured metadata (company, year, section, form type). \textsc{RAGMATE-10K} is publicly available.\footnotemark[\value{footnote}]

We create 30 human-authored templates that instantiate into company- and year-specific questions, covering both general (e.g., business overview) and in-depth (e.g., risk factors) information needs. Excluding Apple filings from evaluation avoids contamination, leaving \( 120 \) test queries.  

Ground-truth answers are generated by constraining a language model to use only chunks from the target filing. The model must cite the supporting chunks, providing supervision for both retrieval accuracy and answer grounding.  

\subsection{Implementation Details}
We isolate metadata design effects by using a frozen text encoder \( f_{\theta} \) and a fixed retrieval pipeline. Each retrieval unit is a pair \( (m_i, c_i) \), where \( c_i \) is a chunk of document text and \( m_i \) is a flat key–value metadata dictionary.  

We evaluate two representative embedding models: OpenAI’s \texttt{text-embeddi-} \texttt{ng-3-small} (dimension 1536) \cite{openai_embedding} and BAAI’s \texttt{bge-m3} (dimension 1024) \cite{noauthor_baaibge-m3_2025}, a strong open-source retriever optimized for multilingual and cross-domain retrieval. Both encoders are used in frozen form without fine-tuning.  

Metadata \( m_i \) is represented as a flat key–value dictionary, independent of the chunk text. The fields are serialized into a fixed-order header for text-based variants, and passed verbatim to the metadata encoder for dual-encoder variants.  
\begin{table}[t]
\caption{Flat metadata schema used in all experiments.}
\label{tab:metadata_schema}
\centering
\small
\begin{tabularx}{0.8\textwidth}{lll}
\toprule
\textbf{Field} & \textbf{Description} & \textbf{Example} \\
\midrule
\texttt{company\_name} & Filing entity name & Alphabet Inc. \\
\texttt{form\_type} & SEC form type & 10-K \\
\texttt{section} & Document section heading & Item 1 - BUSINESS \\
\texttt{fiscal\_year\_end} & Fiscal year end date & 12-31 \\
\texttt{period\_of\_report} & Reporting period close date & 2023-12-31 \\
\texttt{filed\_date} & SEC filing submission date & 2024-01-31 \\
\texttt{exchange\_listings} & Public exchange(s) listed on & [NYSE] \\
\texttt{SIC\_code} & Industry classification & COMPUTER \\
\bottomrule
\end{tabularx}
\end{table}
\begin{figure}[t]
    \centering
    \begin{subfigure}{0.48\linewidth}
        \centering
        \includegraphics[width=\linewidth]{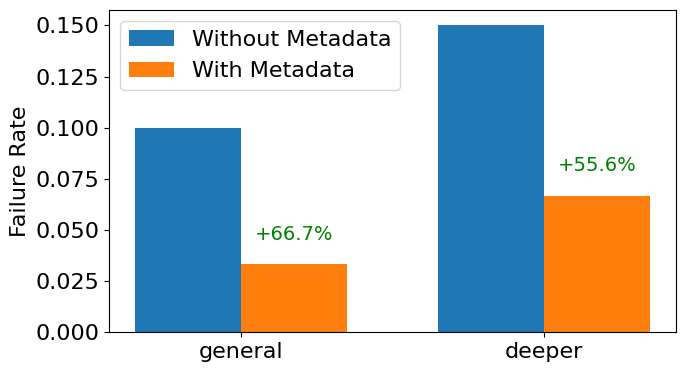}
        \caption{Retrieval failure rate by category.}
    \end{subfigure}
    \hfill
    \begin{subfigure}{0.48\linewidth}
        \centering
        \includegraphics[width=\linewidth]{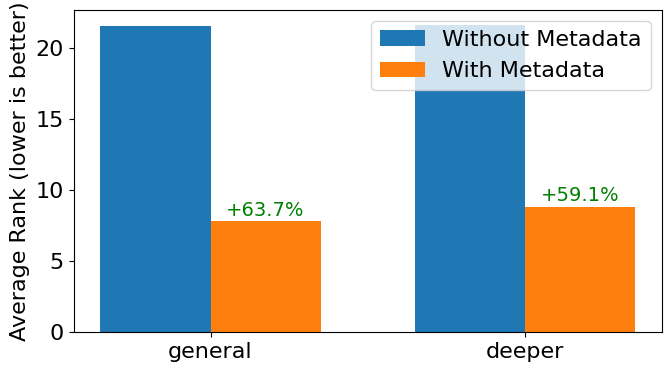}
        \caption{Average rank of first match.}
    \end{subfigure}
\caption{Comparative retrieval performance vs.\ plain baseline across query types using the Dual Encoder Unified Embedding approach.}
    \label{fig:uni_findings_2}
\end{figure}
\begin{figure}[t]
    \centering
    \begin{subfigure}{0.48\linewidth}
        \centering
        \includegraphics[width=\linewidth]{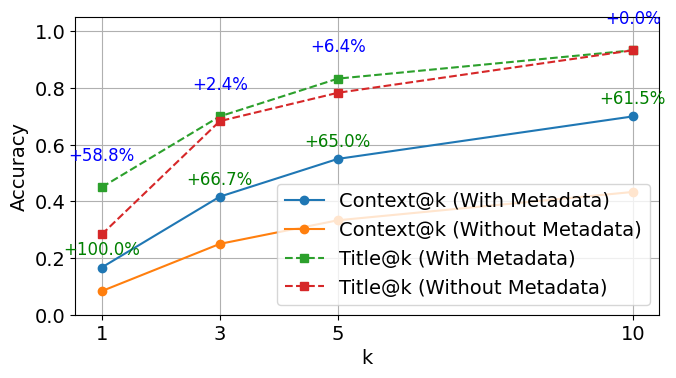}
        \caption{Context@K and Title@K (general questions).}
    \end{subfigure}
    \hfill
    \begin{subfigure}{0.48\linewidth}
        \centering
        \includegraphics[width=\linewidth]{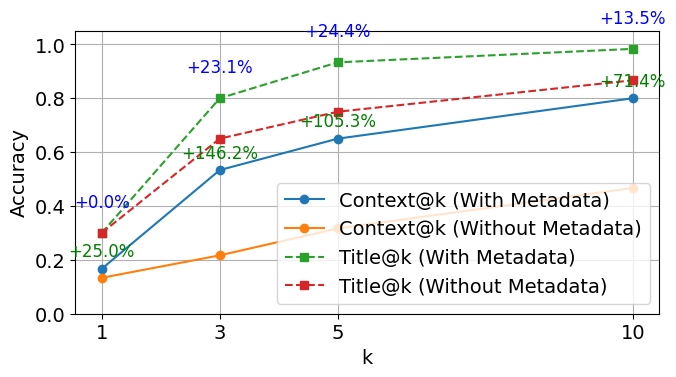}
        \caption{Context@K and Title@K (in-depth questions).}
    \end{subfigure}
    \begin{subfigure}{0.48\linewidth}
        \centering
        \includegraphics[width=\linewidth]{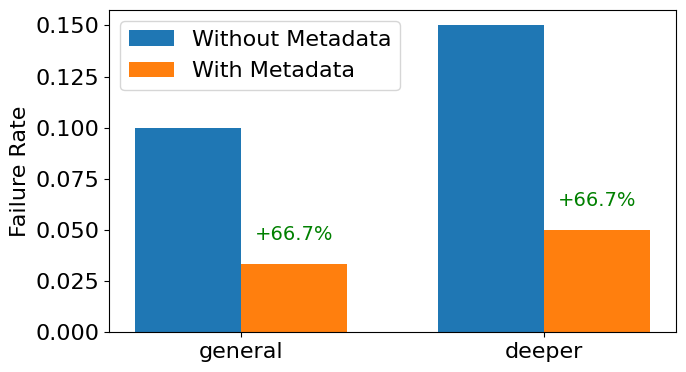}
        \caption{Retrieval failure rate by category.}
    \end{subfigure}
    \hfill
    \begin{subfigure}{0.48\linewidth}
        \centering
        \includegraphics[width=\linewidth]{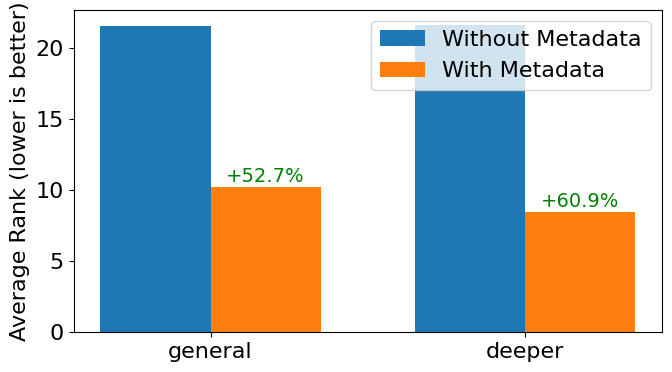}
        \caption{Average rank of first match.}
    \end{subfigure}
    \caption{Comparative retrieval performance vs.\ plain baseline across query types using Metadata as Text (Prefix)}
    \label{fig:mat_findings}
\end{figure}
We evaluate retrieval quality using cosine similarity with top-$K$ search, varying $K \in \{1,\dots,10\}$. Performance is measured against ground-truth supporting chunks using four metrics:  
\begin{itemize}
    \item \textbf{Context@K}: whether at least one retrieved chunk within the top $K$ supports the ground-truth answer.  
    \item \textbf{Title@K}: whether at least one top-$K$ chunk comes from the correct document (company and year).  
    \item \textbf{Average Matched Rank}: the mean rank position of the highest-ranked supporting chunk (lower is better).  
    \item \textbf{Retrieval Failure Rate}: the proportion of queries for which no relevant chunk is retrieved (lower is better).  
\end{itemize}

\section{Findings}
\label{sec:findings}
Our experiments show that metadata-aware retrieval consistently improves over a plain content-only baseline. Among the strategies, two stand out: prefix-based metadata-as-text (MaT) and dual-encoder unified embeddings.  

Unified embeddings emerge as the most effective and practical approach. By fusing metadata and content vectors into a single index, they achieve accuracy that matches or surpasses prefixing while offering clear advantages in index maintenance and serving. This makes unified embeddings a strong candidate for deployment in real-world RAG systems where metadata evolves over time.  

At the same time, metadata-as-text remains a simple and high-performing baseline. Direct concatenation reliably boosts retrieval accuracy and requires no architectural changes or additional infrastructure, making it appealing as a training-free baseline.  

\begin{figure}[H]
    \centering
    \includegraphics[width=0.8\textwidth]{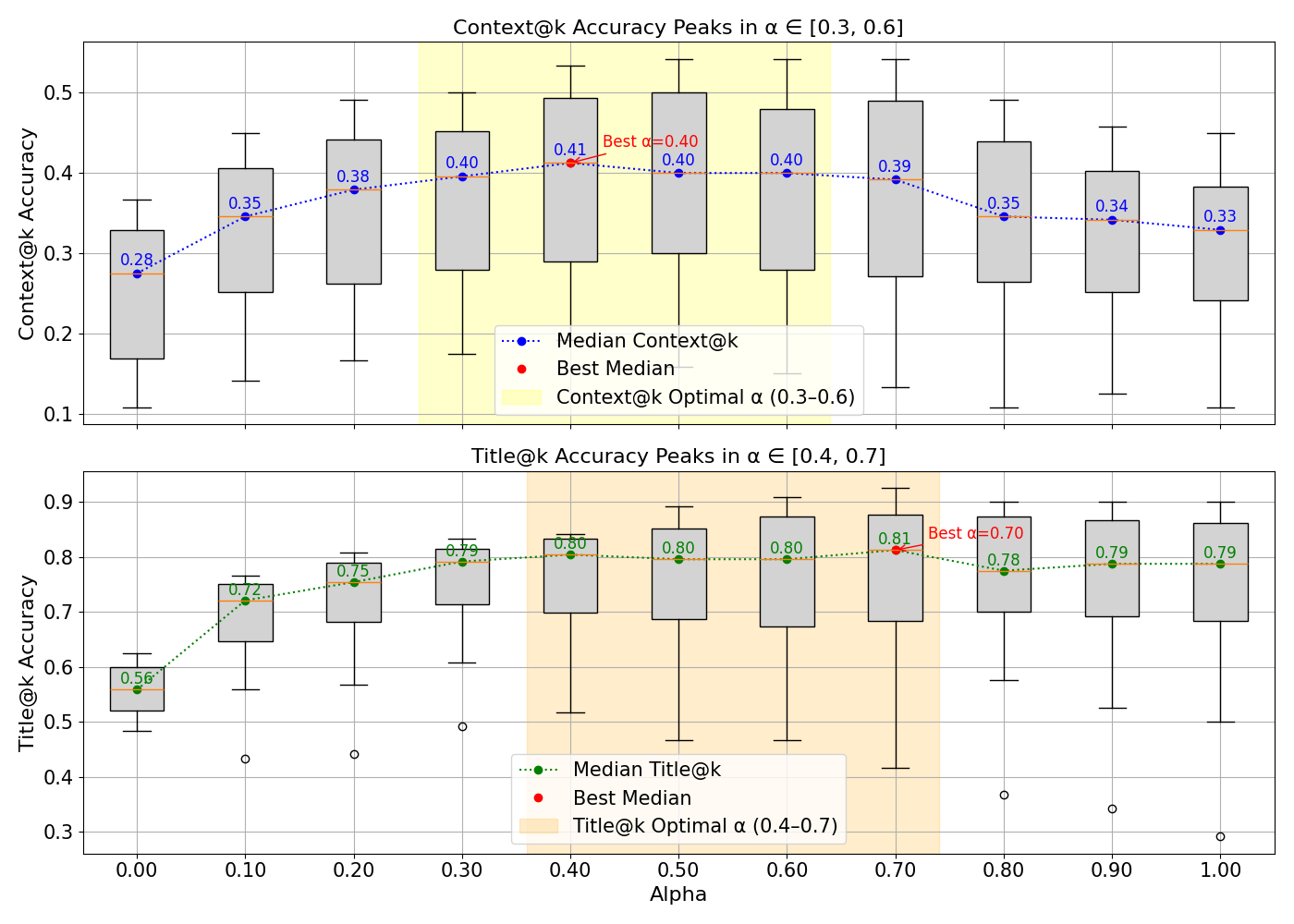}
    \caption{Retrieval performance across metadata weight \( \alpha \) for Dual encoder with late-fusion scoring. Metadata improves results when moderately weighted; full reliance on either content or metadata reduces performance.}
    \label{fig:alpha_sweep_comparison}
\end{figure}

Late-fusion dual encoders and metadata-aware query reformulation trail these methods. Sweeping the fusion weight \( \alpha \in \{0.0, 0.1, \dots, 1.0\} \) shows moderate values (\(\alpha \approx 0.3{-}0.6\)) work best, confirming that metadata should complement, not dominate, content signals. In practice, late fusion offers little advantage: unified embeddings provide similar balance within a single index, avoiding extra lookups and latency. Still, the \(\alpha\)-sweep is useful diagnostically to gauge metadata’s marginal value (see Figure~\ref{fig:alpha_sweep_comparison}).
\begin{table}[t]
\centering
\caption{Retrieval performance summary at cutoff $k{=}5$ across two embedding models. 
Dual encoders are reported with $\alpha=0.5$.}
\label{tab:retrieval_summary}
\begin{subtable}{\textwidth}
\centering
\caption{OpenAI \texttt{text-embedding-3-small}}
\resizebox{\textwidth}{!}{%
\begin{tabular}{lrrrr|rrrr}
\toprule
\multirow{2}{*}{\textbf{Method}} & 
\multicolumn{4}{c|}{\textbf{General}} & 
\multicolumn{4}{c}{\textbf{Deeper}} \\
& \textbf{Context@5 $\uparrow$} & \textbf{Title@5 $\uparrow$} & \textbf{Avg Rank $\downarrow$} & \textbf{Failure $\downarrow$} 
& \textbf{Context@5 $\uparrow$} & \textbf{Title@5 $\uparrow$} & \textbf{Avg Rank $\downarrow$} & \textbf{Failure $\downarrow$} \\
\midrule
No Metadata & 33.33 & 78.33 & 21.61 & 10.00 & 31.67 & 75.00 & 21.63 & 15.00 \\
Meta-Suffix & 48.33 & 88.33 & 11.49 & 1.67 & 53.33 & 86.67 & 10.52 & 3.33 \\
Dual(late fusion) & 36.67 & 78.33 & 10.80 & 15.00 & 38.33 & 78.33 & 15.43 & 15.00 \\
Dual(Reformulated) & 41.38 & 82.76 & 11.04 & 22.41 & 38.60 & 78.95 & 13.89 & 19.30 \\
\textbf{Meta-Prefix} & \textbf{55.00} & \textbf{83.33} & \textbf{10.22} & \textbf{3.33} & \textbf{65.00} & \textbf{93.33} & \textbf{8.46} & \textbf{5.00} \\
\textbf{Dual(Unified)} & \textbf{63.33} & \textbf{88.33} & \textbf{7.84} & \textbf{3.33} & \textbf{60.00} & \textbf{83.33} & \textbf{8.84} & \textbf{6.67} \\
\bottomrule
\end{tabular}%
}
\end{subtable}
\begin{subtable}{\textwidth}
\centering
\caption{BAAI \texttt{bge-m3}}
\resizebox{\textwidth}{!}{%
\begin{tabular}{lrrrr|rrrr}
\toprule
\multirow{2}{*}{\textbf{Method}} & 
\multicolumn{4}{c|}{\textbf{General}} & 
\multicolumn{4}{c}{\textbf{Deeper}} \\
& \textbf{Context@5 $\uparrow$} & \textbf{Title@5 $\uparrow$} & \textbf{Avg Rank $\downarrow$} & \textbf{Failure $\downarrow$} 
& \textbf{Context@5 $\uparrow$} & \textbf{Title@5 $\uparrow$} & \textbf{Avg Rank $\downarrow$} & \textbf{Failure $\downarrow$} \\
\midrule
No Metadata & 26.67 & 93.33 & 16.45 & 30.00 & 33.33 & 95.00 & 15.95 & 31.67 \\
Meta-Suffix & 33.33 & 95.00 & 17.93 & 6.67 & 45.00 & 95.00 & 11.00 & 11.67 \\
Dual(late fusion) & 30.00 & 88.33 & 13.67 & 35.00 & 31.67 & 91.67 & 19.37 & 31.67 \\
Dual(reformulated) & 25.86 & 86.21 & 15.74 & 34.48 & 26.32 & 94.74 & 16.94 & 40.35 \\
\textbf{Meta-Prefix} & \textbf{61.67} & \textbf{96.67} & \textbf{6.97} & \textbf{3.33} & \textbf{58.33} & \textbf{95.00} & \textbf{10.98} & \textbf{5.00} \\
\textbf{Dual(Unified)} & \textbf{55.00} & \textbf{93.33} & \textbf{10.73} & \textbf{8.33} & \textbf{51.67} & \textbf{95.00} & \textbf{10.09} & \textbf{10.00} \\
\bottomrule
\end{tabular}%
}
\end{subtable}
\end{table}
Across both embedding models, OpenAI’s \texttt{text-embedding-3-small} (1536-dim) and BAAI’s \texttt{bge-m3} (1024-dim), the metadata-enriched strategies consistently outperform the plain baseline. The gains are most pronounced for complex queries that require company- or section-level disambiguation. Figures~\ref{fig:uni_findings_1},\ref{fig:uni_findings_2} and~\ref{fig:mat_findings} visualize these comparative performance trends across query types.

Table~\ref{tab:retrieval_summary} reports representative metrics at $k{=}5$, providing a compact numerical view of the performance differences across methods. These headline results set the stage for deeper analyses of trade-offs and ablations in the following sections.

\subsection{Analysis of Embedding Space}
\label{sec:embedding_analysis}
We test Propositions~\ref{prop:intra}--\ref{prop:variance} by computing pairwise cosine similarities under both metadata-as-text (MaT) and unified embeddings (Eq.~\ref{eq:sumfusion_doc}). For each pair of chunks we form two similarities, plain \(\cos(\mathbf{e}_i,\mathbf{e}_j)\) and metadata-enriched \(\cos(\tilde{\mathbf{e}}^{\text{MaT}}_i,\tilde{\mathbf{e}}^{\text{MaT}}_j)\) or \(\cos(\tilde{\mathbf{e}}^{\text{Unif}}_i,\tilde{\mathbf{e}}^{\text{Unif}}_j)\). We stratify pairs into Same Company \& Year (positive) and Different (negative), and quantify separation.
\begin{figure}[t!]
    \centering
    \includegraphics[width=0.8\linewidth]{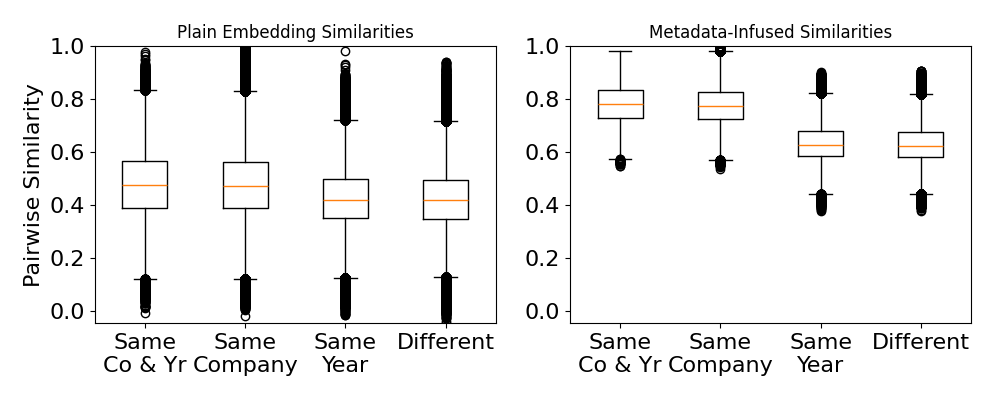}
    \caption{Embedding space analysis with unified embeddings (grouped intra-/inter-document similarities)}
    \label{fig:embedding_analysis}
\end{figure}
\begin{table}[b]
\centering
\caption{Separation between Same Company \& Year (pos) and Different (neg) pairs, computed from pairwise cosine similarities. We report plain baseline values and relative improvements (\(\Delta\)) for unified and prefix embeddings. Arrows indicate whether higher (\(\uparrow\)) or lower (\(\downarrow\)) values are better.}
\label{tab:pairwise_separation}
\small
\begin{tabular}{l r rr}
\toprule
\textbf{Metric} & \textbf{Plain} & \(\Delta\) \textbf{Dual(Unified)} & \(\Delta\) \textbf{Meta-Prefix} \\
\midrule
Mean margin (\(\uparrow\)) & 0.054 & +0.098 & \textbf{+0.101} \\
Cohen’s \(d\) (\(\uparrow\)) & 0.450 & \textbf{+1.800} & +1.440 \\
Fisher score \(F\) (\(\uparrow\)) & 0.102 & \textbf{+2.425} & +1.692 \\
AUC (\(\uparrow\)) & 0.625 & \textbf{+0.315} & +0.283 \\
KS distance (\(\uparrow\)) & 0.190 & \textbf{+0.520} & +0.461 \\
Tail mass \(\Pr[\cos \ge 0.8]\), pos (\(\uparrow\)) & 0.007 & \textbf{+0.407} & +0.108 \\
Tail mass \(\Pr[\cos \ge 0.8]\), neg (\(\downarrow\)) & 0.000 & +0.006 & \textbf{+0.000} \\
\bottomrule
\end{tabular}
\end{table}

Below we report results for the dual-encoder unified embeddings; the MaT variant shows similar trends. Metadata increases similarity across all strata but especially for positive pairs, widening the gap. For example, the mean margin between positives and negatives triples (0.054$\rightarrow$0.152), Cohen’s \(d\) grows from a small effect (0.45) to a very large one (2.25), and AUC rises from 0.63 to 0.94. As Table~\ref{tab:pairwise_separation} shows, both metadata strategies yield large gains over plain embeddings, with unified embeddings outperforming prefixing on majority of all metrics while being easier to maintain.

\section{Impact of Metadata Fields: Ablation Study}
\label{sec:context}
Not all metadata fields contribute equally to retrieval performance. Here we focus on two types of signals: global identifiers such as company and year, and local context from section titles.  

Chunks extracted from long, repetitive documents like SEC filings often lose their contextual anchor. For example, the phrase ``we believe our strategy is working'' has very different implications in a ``Risk Factors'' section versus in ``Management’s Discussion and Analysis.'' We distinguish two types of context: i) global, provided by fields such as company and year that identify the document as a whole, and ii) local, provided by fields such as section titles that situate a chunk within its broader structure. To test their contributions, we compare four conditions: i) No Metadata (Baseline): plain chunk text without metadata; ii) Full Metadata: all fields, including section titles; iii) w/o section: Full metadata except section titles. iv) w/o company and year: Full metadata except company and year. All embeddings use OpenAI’s \texttt{text-embedding-3-small} model with the MaT formulation, and retrieval is evaluated with the metrics from Section~\ref{sec:methods}, focusing on Context@K and Title@K.

\begin{figure}[tbp]
    \centering
    \begin{subfigure}{0.48\linewidth}
        \centering
        \includegraphics[width=\linewidth]{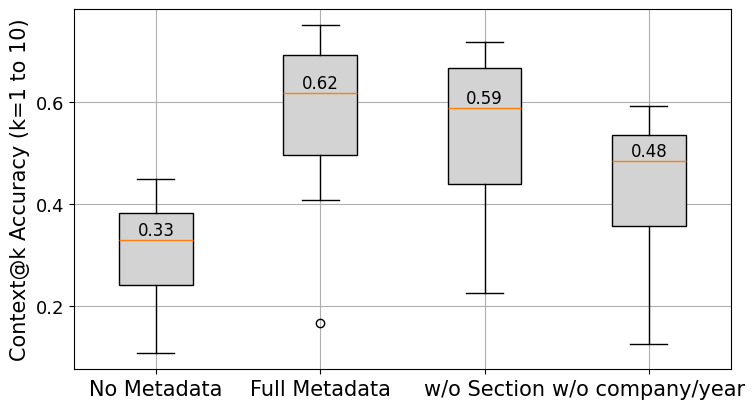}
        \caption{Context@K accuracy ($k=1$ to $10$)}
        \label{fig:context_ablation}
    \end{subfigure}
    \hfill
    \begin{subfigure}{0.48\linewidth}
        \centering
        \includegraphics[width=\linewidth]{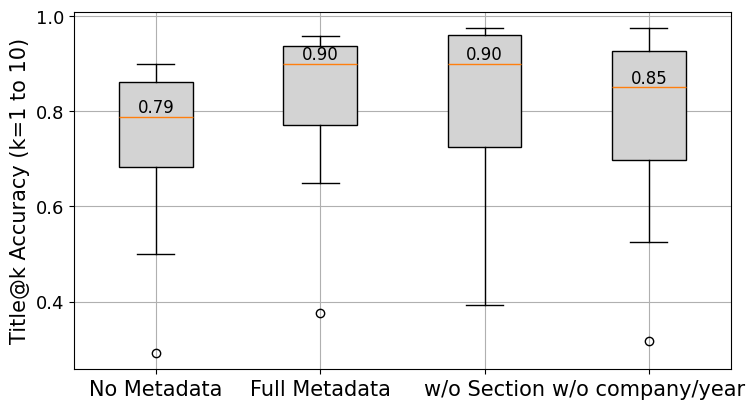}
        \caption{Title@K accuracy ($k=1$ to $10$).}
        \label{fig:title_ablation}
    \end{subfigure}
    \caption{Impact of metadata ablations on retrieval performance}
    \label{fig:ablation_combined}
\end{figure}

Figures~\ref{fig:context_ablation} and~\ref{fig:title_ablation} report the results. Company and year provide the strongest disambiguating signal: removing them reduces both Title@K and Context@K. In contrast, removing section titles yields only a modest drop in Context@K with no effect on Title@K, suggesting that global identifiers drive document-level accuracy while section cues primarily aid chunk-level localization.

\section{Conclusion, Limitations, and Future Work}
We revisited the role of metadata as a first-class retrieval signal in RAG, using SEC filings as a testbed. Across two embedding models, we found that embedding metadata alongside content consistently improves retrieval. Simple prefixing of metadata is strong, but a dual-encoder with unified embeddings matches or exceeds its accuracy while being easier to maintain. Field-level ablations show that company and year act as strong disambiguators, while section titles are only modestly useful, and embedding-space analysis reveals that metadata improves geometry by tightening intra-document similarity and reducing cross-document confusion.  

We also acknowledge limitations. Our study focuses on SEC Form 10-K filings, a deliberately challenging stress-test corpus characterized by rigid templates, heavy lexical repetition, and subtle document-level distinctions that frequently confound semantic-only retrieval. While absolute gains may vary across domains, these worst-case conditions allow us to isolate the effect of metadata integration, and we expect the relative benefits of treating metadata as a first-class retrieval signal, particularly via unified embeddings, to generalize to other structured corpora such as scientific articles, legal records, and technical manuals.
Ground-truth answers were generated in a semi-supervised manner, a limitation that nevertheless reflects the common practice of using LLMs as judges or oracles \cite{Zheng2023JudgingLW,Fu2023GPTScoreEA,li2023generative_judge,Hackl_2023,huang2024empirical_judge}. In addition, we evaluate frozen encoders without exploring fine-tuning or end-to-end training.  

Overall, our results suggest that effective metadata integration does not require complex architectures: concatenation or unified fusion already offer a strong balance of accuracy and practicality. Future directions include adaptive weighting, richer metadata modalities (tables, figures), and human evaluation of downstream generation quality.  
\begin{credits}
\subsubsection{\discintname}
The authors have no competing interests to declare that are relevant to the content of this article.
\end{credits}
\bibliographystyle{splncs04}
\bibliography{ref}
\end{document}